\documentstyle[14lomcon,cite,amsmath,amsfonts,epsfig,graphics]{article}

\begin{document}

\title{SEMI-INCLUSIVE PION ELECTROPRODUCTION WITH CLAS}

\author{ Mikhail Osipenko \footnote{e-mail: osipenko@ge.infn.it}}

\address{INFN, sez. di Genova, 16146 Genova, Italy, \\
SINP, Moscow State University, 119991 Moscow, Russia}


\maketitle\abstracts{
Measurement of the semi-inclusive $\pi^+$ electroproduction off the proton,
performed with CLAS detector at Jefferson Lab, has been presented.
The obtained fully-differential cross sections,
including the azimuthal angle between hadronic and leptonic planes, $\phi$,
allowed us to separate the $\phi$-dependent terms.
While, the $\phi$-independent part of the cross section
was found to be in good agreement with current fragmentation pQCD calculations.}

The semi-inclusive electroproduction of hadrons is an important tool
for studying the nucleon structure in the perturbative Quantum Chromodynamics (pQCD) framework
at medium energies. Indeed, the detection of a hadron produced by the struck quark
or by nucleon spectator fragments provides an information about the orbital momentum
of the quark in the initial state. Meanwhile, the undetected hadronic system
allows to apply the optical theorem, reducing the energy necessary for the convergence towards
basic pQCD processes.

CLAS has measured the semi-inclusive electroproduction of $\pi^+$ on the proton
at the beam energy of 6 GeV. The measurement span over a wide, continuous 5-dimensional
domain, which allows for a detailed study of $\phi$ and $p_T$ behaviors.
In this article we will focus on one particular aspect of the obtained results.
This aspect deals with the comparison of the data, integrated in $\phi$
and transverse momentum $p_T$, expressed in terms of the structure function $H_2$:
\begin{equation}\label{eq:h2}
\frac{d^3 \sigma}{dx dQ^2 dz} = \frac{4\pi\alpha^2}{x Q^4}
\Biggl [x y^2 H_1(x,z,Q^2) + \Bigl(1-y-\frac{M^2x^2y^2}{Q^2}\Bigr) H_2(x,z,Q^2) \Biggr ] ~~~,
\end{equation}
with pQCD calculations~\cite{Furmanski}.
These calculations describe the structure function $H_2$
as the convolution of the parton density function $f(x,Q^2)$ obtained in inclusive processes
and the parton fragmentation function $D^h(z,Q^2)$ measured in $e^+e^-$ collisions:
\begin{equation}\label{eq:h2_pqcd}
H_2(x,z,Q^2)=\sum_i e_i^2 x f_i(x,Q^2) \otimes D_i^h(z,Q^2) ~,
\end{equation}
\noindent where the sum runs over quark flavors $i$ and $e_i$ is the charge of $i$th flavor quark
(we neglect the gluon contribution here).

In particular, we are interested in the difference between data-theory comparisons
made using the structure function $H_2$ and structure function ratio $H_2/F_2$.
The latter ratio, where the $F_2$ is the inclusive structure function,
represents in DIS limit the widely used multiplicity observable,
and can be calculated in pQCD as following:
\begin{equation}\label{eq:h2f2_pqcd}
\frac{H_2(x,z,Q^2)}{F_2(x,Q^2)}=\frac{\sum_i e_i^2 x f_i(x,Q^2) \otimes D_i^h(z,Q^2)}{\sum_i e_i^2 x f_i(x,Q^2)} ~.
\end{equation}
\noindent 
Given the relatively low beam energy of Jefferson Lab one may expect
a manifestation of visible deviations from pQCD calculations at the low-$Q^2$ end
of the covered interval. Such deviations should indicate the contribution
of higher twists in the semi-inclusive electroproduction. The detailed
calculations of these higher twists are not available due to their complexity.
However, one may phenomenologically divide them in two types:
Initial State Interactions (ISI) of the current quark
and Final State Interactions (FSI) of struck quark or produced hadron.
If the dominant contribution to the total higher twist term would
be due to ISI, one could expect a partial cancellation of them in $H_2/F_2$ ratio.
Hence, it is possible that $H_2/F_2$ ratio agrees with pQCD calculations better
that the $H_2$ structure function alone.

\begin{figure}[!h]
\begin{center}
\includegraphics[bb=2cm 13cm 22cm 23cm, scale=0.4]{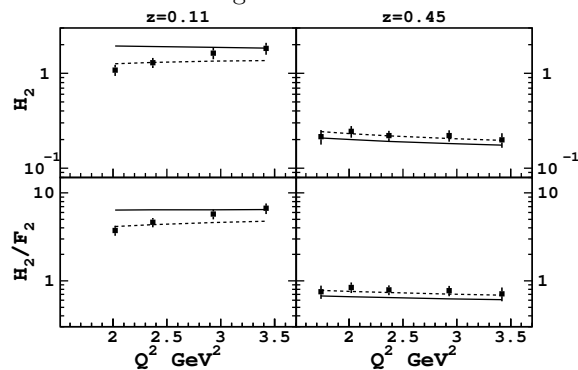}
\caption{The $Q^2$-evolution of the data on structure function $H_2$ and ratio $H_2/F_2$
at $x=0.34$ and two values of $z$,
in comparison to pQCD calculations: LO - solid line, NLO - dashed line.
The calculations use CTEQ 5 parton distributions~\cite{CTEQ},
and Kretzer fragmentation functions~\cite{Kretzer}.
The error bars give statistical and systematic uncertainties combined in quadrature.}
\label{fig:cmp1}
\end{center}
\end{figure}

In Fig.~\ref{fig:cmp1} comparisons of the measured structure function $H_2$ and $H_2/F_2$ ratio
to LO and NLO pQCD calculations are shown for two values of $z$.
As one can see, the NLO calculations describe very well the data at $z=0.45$
for both observables, while at $z=0.11$ some deviation in the $Q^2$-slope is evident.
This deviation can be due to the higher twist contribution.

\begin{figure}[!h]
\begin{center}
\includegraphics[bb=2cm 13cm 22cm 23cm, scale=0.4]{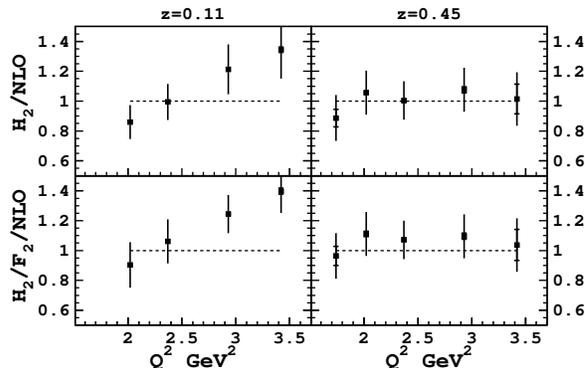}
\caption{Same as Fig.~\ref{fig:cmp1} except with the ratio of $H_2$ and $H_2/F_2$ data
to the NLO calculations. The dashed line indicates the unity. The inner error bars
(mostly smaller than the symbol size) give statistical uncertainties.}
\label{fig:cmp2}
\end{center}
\end{figure}

In Fig.~\ref{fig:cmp2} the same comparison is shown for the ratio of the data
over NLO calculations. In the amplified scale the deviation of the low-$z$ data
from the expected NLO evolution can be quantified. Both $H_2$ and $H_2/F_2$
comparisons show the deviation rising with $Q^2$ from -10\% up to 40\%.
The difference between $H_2$ and $H_2/F_2$ ratios results in a few percent overall shift,
well below the systematic uncertainties of the measurement (about 15\% in average)
and the theory.

Summarizing, the use of $H_2$ structure function or $H_2/F_2$ ratio
in the comparison of the experimental data to pQCD calculations is equivalent in JLab energy domain.
If the deviations of the data from NLO pQCD calculations, observed at low-$z$,
are due to the higher twist contribution, they are probably related to the FSI mechanism.
However, given the low-$z$ values at which the difference is observed,
an alternative explanation due to the mixing between current and target
fragmentation evolutions is more likely.
The pQCD calculations described above are due to the current fragmentation only.
The target fragmentation mechanism~\cite{Trentadue}, expected to play role in the low-$z$ domain,
is still poorly established. In particular, the corresponding partonic functions,
fracture functions for the pions are completely unknown.
This encourages further pQCD studies of semi-inclusive reactions at Jefferson Lab.

\section*{Acknowledgments}

Author would like to express his gratitude to Prof. A.Kataev for kind invitation
and Prof.B.S.Ishkhanov for the help with the present proceeding.


\section*{References}
\begin{thebibliography}{99}

\bibitem{sidis_pip} M.Osipenko {\it et al.}, {\it Phys.Rev.} {\bf D} 80, 032004 (2009).

\bibitem{CTEQ} H.L.Lai {\it et al.}, {\it Eur.Phys.J.} {\bf C} 12, 375 (2000).

\bibitem{Kretzer} S.Kretzer, {\it et al.}, {\it Phys.Rev.} {\bf D} 62, 054001 (2000).

\bibitem{Furmanski} W.Furmanski and R.Petronzio, {\it Z.Phys.} {\bf C} 11, 293 (1982).

\bibitem{Trentadue} L.Trentadue and G.Veneziano, {\it Phys.Lett.} {\bf B} 323, 201 (1994).

\end{thebibliography}
\end{document}